# Ultra-Low-Temperature Reactions of Carbon Atoms with Hydrogen Molecules

S. A. Krasnokutski[1], M. Kuhn[2], M. Renzler[2], C. Jäger[1], Th. Henning[1], P. Scheier[2]

[1] *Laboratory Astrophysics Group of the Max Planck Institute for Astronomy at the Friedrich Schiller University Jena, Institute of Solid State Physics, Helmholtzweg 3, D-07743 Jena, Germany*

[2] *Institute for Ion Physics and Applied Physics, Technikerstr. 25, A-6020 Innsbruck, Austria.*

**Abstract:** The reactions of carbon atoms with dihydrogen have been investigated in liquid helium droplets at $T = 0.37$ K. A calorimetric technique was applied to monitor the energy released in the reaction. The barrierless reaction between a single carbon atom and a single dihydrogen molecule was detected. Reactions between dihydrogen clusters and carbon atoms have been studied by high-resolution mass spectrometry. The formation of hydrocarbon cations of the type $C_mH_n^+$, with $m = 1 - 4$ and $n = 1 - 15$ was observed. With enhanced concentration of dihydrogen, the mass spectra demonstrated the main "magic" peak assigned to the $CH_5^+$ cation. A simple formation pathway and the high stability of this cation suggest its high abundance in the interstellar medium.

## 1. INTRODUCTION

According to different estimations, a large portion of carbon in the interstellar medium (ISM) exists in the form of an atomic gas (Snow & Witt 1995), and the most abundant molecule present in the ISM is dihydrogen. Despite its fundamental importance for astrochemistry, up to now, not much is known about chemical reactions between carbon atoms and dihydrogen. In the gas phase, the endothermic reaction $C + H_2 \rightarrow CH + H$ takes place (Dean et al. 1991). The reverse reaction $CH + H \rightarrow C + H_2$ is also possible (Becker et al. 1989). Both of these reactions proceed via the HCH reactive intermediate. Stabilization of this intermediate radical becomes possible in the presence of a third body $C + H_2 + M \rightarrow CH_2 + M$, for example, when the reaction proceeds on the surface of dust grains, or via three-body collisions in the gas phase. Such three-body collision reactions were studied experimentally at room temperature in the gas phase by monitoring the decay of the carbon atom abundancies (Husain & Kirsch 1971, Husain & Young 1975, Martinotti et al. 1968). A low reaction rate $k = 6.9 \times 10^{-32}$ cm$^6$ molecule$^{-2}$ s$^{-1}$ of this reaction was found, but no temperature dependent studies were performed (Husain & Young 1975). The reverse reaction $CH + H \rightarrow C + H_2$ was studied in the high-temperature range of 1504 – 2042 K. A relatively strong temperature dependence of the reaction rate was found (Dean et al. 1991, Dean & Hanson 1992). In the same study, no temperature dependence for the reaction $C + H_2 \rightarrow CH + H$ was observed. However, the rate of this reaction was considerably higher than that measured at room temperature (Becker et al. 1989). These results suggest the presence of an energy barrier between $H_2 + C$, $CH_2$, HCH, and $CH + H$ states, meaning extremely low reaction rates for both reaction directions at low temperatures. However, the prediction of the reaction rates at low temperatures, based on the high temperature results, is often an origin of large errors. At the moment, there are no experimental studies of these reactions at low temperatures, while all quantum chemical computations performed do not find any notable energy barrier in the reaction pathways (Bussery-Honvault et al. 2005, Harding et al. 1993, Lin & Guo 2004). Additionally, there is a lack of data for the reaction of HCH molecules with dihydrogen (Ge et al. 2010).

In this article, we investigated the reaction of carbon atoms with dihydrogen inside superfluid helium nanodroplets. All species picked up by the droplets adopt their temperature ($T = 0.37$ K) on a subnanosecond timescale (Grebenev et al. 1998, Toennies & Vilesov 2004). Therefore, all reactants equilibrate to this well-known temperature before they meet and react with each other. The liquid helium absorbs the reaction energy, allowing associative reactions of the type $A + B \rightarrow AB$. Thus, reactions inside liquid helium droplets are close analogs of the reactions on the surface of dust grains. Due to the nanoscale size of helium droplets, it is possible to dope them by a single atom or a molecule of each reactant. Mass spectrometry is a convenient tool to monitor the accomplishment of chemical reactions and analyze the reaction products (Denifl et al. 2009, Schöbel et al. 2011, Shepperson et al. 2011, Thaler et al. 2015).

Additionally, the helium droplet isolation technique allows the use of a unique detection method. The energy released in the reaction leads to the evaporation of a given number of helium atoms from the surface of the helium droplet. Therefore, by measuring the helium droplet size before and after the incorporation of reactants, it is possible to obtain qualitative (Krasnokutski & Huisken 2010a, 2011) and, in some cases, even quantitative (Krasnokutski et al. 2014) information on the energy released during the reaction.

## 2. EXPERIMENTAL DETAILS

The experiments have been carried out in two different helium droplet beam apparatus reported earlier (Krasnokutski et al. 2005, Schöbel et al. 2010). Both experimental setups were very similar. Large helium clusters were produced by the supersonic expansion of helium gas (99.9999% purity) at high pressure ($p$ = 20 bar) through a cooled 5 μm diameter pinhole nozzle. The size of helium droplets was varied by adjusting the nozzle temperature (Toennies & Vilesov 2004). After skimming, the helium droplets were sequentially doped with the reactants in two separate pick-up cells. Atomic carbon was produced in a new source developed and described very recently (Krasnokutski & Huisken 2014). Briefly, synthetic graphite (Sigma Aldrich, molecular weight 12.01) or Carbon-$^{13}$C (Sigma Aldrich, 99.2 atom % $^{13}$C) powders were loaded into a tantalum tube, which was wrapped from a 0.05 mm thick tantalum foil (inner diameter: 1.4 mm, length: 20 mm). The tube was not welded. After filling, the ends of the tube were clamped and connected to electrodes allowing the tube to be heated up to 2400 K by an electric current. The source provides an intense flux of low-energy (thermal) carbon atoms without the presence of impurities. To confine the carbon atoms to a well-defined volume, the Ta tube was surrounded by a heat shield with two holes for the He droplet beam. Hydrogen gas (Air Liquide, 99.9995% purity) was introduced to the pick-up region from the outside through a leak valve. Carbon-13 was selected to avoid a close match between masses of carbon atoms and helium clusters.

After having traversed the pick-up regions, the helium droplet beam was introduced into a differentially pumped detector chamber. The setup in Innsbruck was equipped with the high-resolution mass spectrometer and was used for all mass spectrometry experiments present in the article. The setup in Jena was used for all calorimetry measurements.

### 2.1. Mass spectrometry detection

Helium droplets with an estimated mean size of $10^5$ helium atoms were used. Helium droplets were first doped with dihydrogen and later in the second pick-up oven with carbon atoms. After arriving to the next differentially pumped vacuum chamber, the doped helium droplets are continuously ionized by electron impact using 70-eV electrons. The ions are pulse extracted to a commercial time-of-flight mass spectrometer equipped with a reflectron (Tofwerk AG, model HTOF). The implemented two-stage reflectron scheme allows to achieve a high-mass resolution of m/Δm ~ 3500 and the accuracy of the mass calibration better than 0.02 amu. Ions are detected by a microchannel plate detector that operates in an ion-counting mode. The residual gas pressures were $2 \times 10^{-5}$ and $1 \times 10^{-5}$ Pa in the pick-up and detector chambers, respectively. The presented mass spectra are the average of 1750 one-second measurements.

### 2.2. Calorimetry measurements

Helium droplets with an estimated mean size of $1.2 \times 10^4$ helium atoms were used. These droplets were first doped with carbon atoms and later with dihydrogen. The residual gas pressures were $3 \times 10^{-5}$ and $1.5 \times 10^{-7}$ Pa in the pick-up and detector chambers, respectively. The life time of the helium droplet in the experimental setup is only a few milliseconds. Therefore, this pressure is sufficient to keep most of the droplets free from impurities. The total pumping speed of all pumps in the detector chambers was 600 l/s. Helium droplets arriving in the detector chamber are completely evaporated after collisions with the walls of the vacuum chamber. This causes the rise of the pressure in this chamber. The pressure in the detector chamber was recorded with a precision ion gauge (Varian UHV-24). To obtain the helium pressure in that chamber the following procedure was performed. After each depletion measurement, the helium droplet beam was blocked by the shutter in the source chamber. Thirty seconds after closing the shutter, the residual gas pressure was measured and subtracted from the obtained values. Finally, the gas correction factor for helium (0.18) was applied. The depletion of the helium droplet beam flux due to the carbon incorporations were calculated as a ratio of the helium pressures in the detector chamber before and 12 seconds after doping the helium droplets with carbon atoms. This short time interval provides the high accuracy of the obtained depletion values despite the long-term instability of the source.

## 3. RESULTS AND DISCUSSION

### 3.1. Calorimetry measurements

The energy released in the reaction inside helium droplets leads to the evaporation of the droplets and reduction of their sizes. As a result, a smaller amount of helium atoms arrive at the detector chamber. The evaporating helium droplets are the main source of helium in the detector chamber. The gaseous helium is continuously evacuated from this chamber with a constant speed. Therefore, the pressure in the detector chamber shows the flux of the helium droplet beam and the depletion of this beam flux shows the amount of the energy released in the reaction. Figure 1 displays the change of the helium pressure in the detector chamber as a function of time. The depletion peaks are due to the doping of helium droplets with carbon atoms in the first pick-up chamber. In the second pick-up chamber the helium droplets can be doped with another reactant (Ar or $H_2$). For better accuracy, the experiment shown in Figure 1 was repeated several times using different doping efficiencies of carbon atoms and average values for the depletions were calculated. We obtained 21.2%, 20.9%, and 27.5% for the depletion of the beam of helium droplets, when the second pick-up chamber is empty, filled with Ar, and $H_2$, correspondingly. As can be seen, the depletion, which is due to the carbon doping was almost equal when the second pick-up chamber was empty or filled with Ar. Moreover, the depletion during the Ar doping was even a bit lower. The binding energy between Ar clusters and carbon atoms is rather low and cannot be detected in this experiment at this level of sensitivity. During the Ar doping, the evaporation of helium droplets, which is due to the released cluster complex formation energy, is overcompensated by the fact that the carbon doped helium droplets have a lower chance of colliding with Ar atoms in the second pick-up cell. The same effect has already been observed when this calorimetric technique was employed to monitor the formation of van der Waals complexes (Krasnokutski & Huisken 2010a).

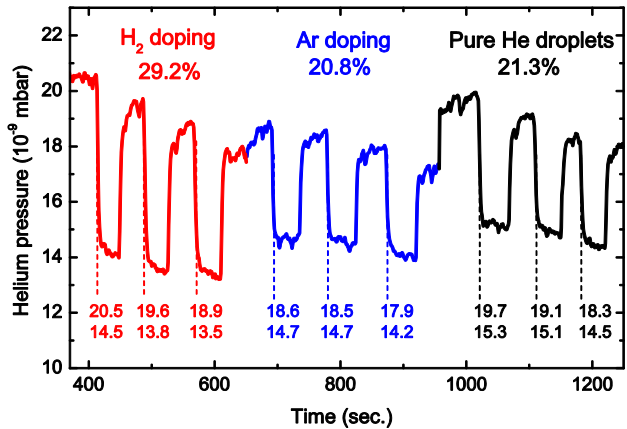

**Figure 1.** Helium pressure in the detector chamber as a function of time. Depletion peaks are due to the doping of the helium droplets with carbon atoms. At the beginning of the experiments, the helium droplets are also doped with dihydrogen, later with argon, and at the end, no additional doping was used. The numbers show the measured helium pressures and the calculated values of the depletion of the helium droplet beam intensity.

The described effect is relatively weak and cannot prevent the detection of a large amount of energy released in the chemical reactions. When the second pick-up oven is filled with dihydrogen, a considerably stronger depletion can be detected. This demonstrates that, in the low-temperature reaction of carbon atoms with dihydrogen, the amount of energy released is much higher than that released in the formation of van der Waals complexes. The heat of evaporation of a single He atom is about 5 $cm^{-1}$ (Toennies & Vilesov 2004). Assuming an extra 7% depletion when Ar gas is replaced with dihydrogen, we can estimate the lower limit of the energy released in the droplets to be about 4200 $cm^{-1}$. The real reaction energy is expected to be much higher because not all helium droplets contain both reactants. Additionally, an ejection of the hot reaction products from the helium droplets, as well as a dissipation of the reaction energy by photon emission, are expected (Krasnokutski & Huisken 2010b).

This estimate is only consistent with the reaction leading to the formation of the HCH molecule because the formation of the $CH_2$ molecule would lead to the release of only a small amount of energy ~600 $cm^{-1}$ (Harding et. al 1993). The high rate of this reaction found at ultra-low temperature, assumes no energy barrier in the reaction pathway. Therefore, our experiment is in line with the results of quantum chemical computations that predict the barrierless reaction $C + H_2 \rightarrow HCH$ (Harding et al. 1993).

In the case of doping of helium droplets by more than a single dihydrogen molecule, the formed HCH molecule can also further react with a $H_2$ molecule leading to the formation of methane. However, the quantum chemical computations predict the presence of a barrier for this reaction (Ge et al. 2010, Lu et al. 2010). Therefore, we expect that this reaction should not proceed inside the helium droplets and that the reactions of carbon atoms are terminated after the formation of HCH molecules.

### 3.2. Mass spectrometry

In the next step, we applied mass spectrometry for the characterization of products of chemical reactions. In this experiment, larger helium droplets (up to $10^5$ He atoms) were used. The droplets were first doped with multiple dihydrogen molecules and, in the second pick up chamber, one or a few carbon atoms were added. We used two different conditions for the doping with dihydrogen. The helium droplets picked up about 10 and 100 dihydrogen molecules on average at low and high doping conditions, correspondingly. Although, even at the high doping conditions, the C/H ratio is much higher than that present in the ISM, it is low enough to saturate the chemistry with hydrogen (i.e. more than five hydrogen atoms per carbon atom). The doped helium droplets were ionized by electron impact in the head of the TOF mass spectrometer.

The collision of a doped helium droplet with an electron, having energy larger than 25 eV, results in the formation of $He^+$. This positive hole migrates within the droplet until it becomes localized either at a dopant or at a helium dimer to form $He_n^+$ (Scheidemann et al. 1993). Therefore, the cations of dopants and helium clusters as well as their complexes are produced.

Figure 2 shows the mass spectra of helium droplets doped with dihydrogen alone and together with carbon atoms. Both mass spectra were recorded one after another using the same experimental conditions. As can be seen, the probability of the charge transfer to the dopant species is quite low. Therefore, the main intensity of the ion signal is detected on the masses of helium clusters. The efficient complex formation between hydrogen and helium leads to the formation of $He_nH_m$ cations. Hydrogen clusters consisting of unbound protons in their nuclei are always a bit heavier than any other elements with the same number of nucleons. Therefore, pure hydrogen clusters or species with the high hydrogen content can be separated from other isobaric ions using high-resolution mass spectrometry. As can be seen from figure 2, switching on the atomic carbon source leads to the reduction of the number of ions of pure hydrogen clusters. At the same time, there are new peaks appearing in the mass spectra. They were assigned to the different hydrocarbon molecules. Practically all hydrocarbon cations $C_mH_n^+$ with $m = 1 - 4$ and $n = 1 - 15$ are produced. The ion intensities on the masses of hydrocarbon products are summarized in Figure 3 for low and high $H_2$ pressures. The data point of $C_2H_2^+$ is missing in both panels because its mass (28.02126 amu) is too close to the mass of $N_2^+$ (28.00615 amu), which is the most abundant species in the residual gas. The other peaks, caused by residual gas ($H_2O$, $OH$, and $CO_2$), do not interfere with our measurements.

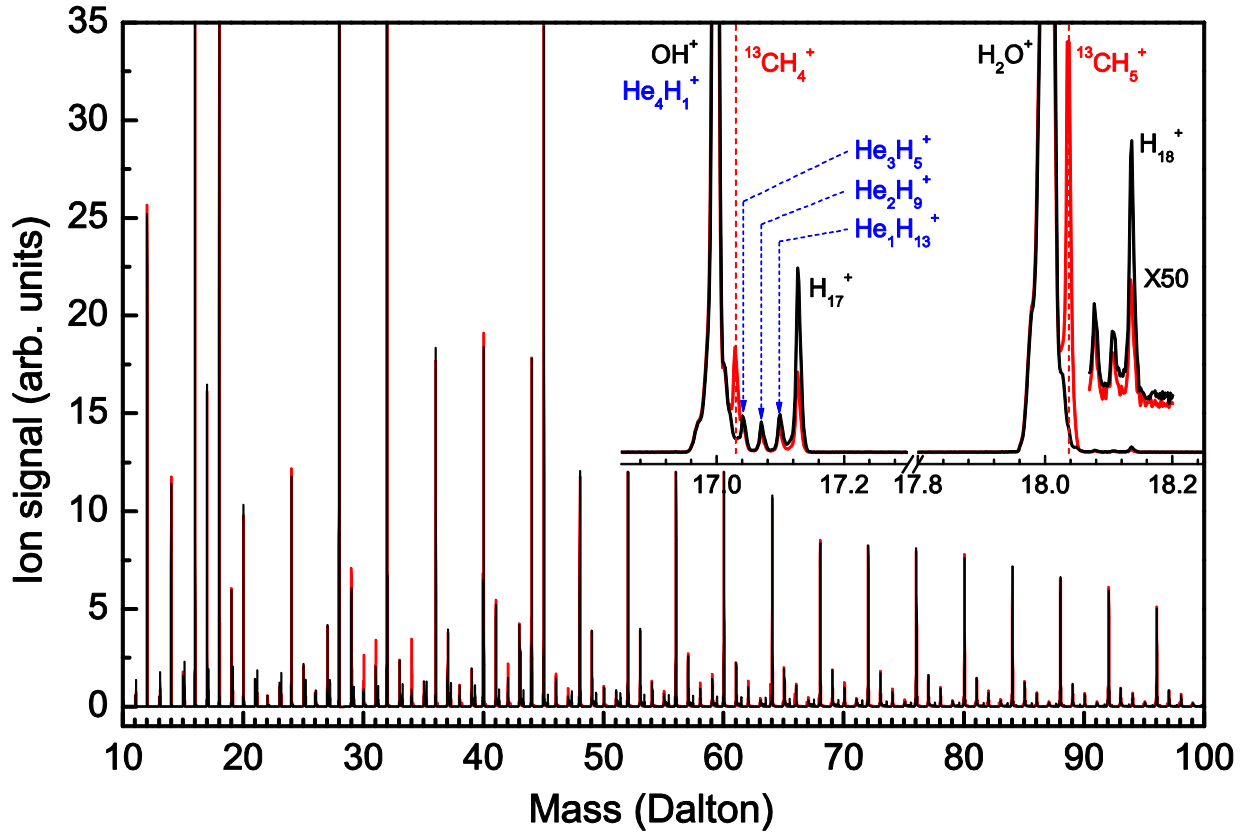

**Figure 2.** Mass spectra of helium droplets doped with $H_2$ molecules only (black) and with $H_2$ molecules and $^{13}C$ atoms together (red). The high doping conditions for dihydrogen were used during recording both mass spectra. Insets show the same mass spectra in a different scale.

As can be seen in the figure, there is only a small cation signal for $n = 1$. This is in line with calorimetry measurements showing that the HCH molecule is formed before the ionization. Therefore, we expect that larger hydrocarbon cations are presumably produced due to ion-molecular reactions after the electron impact ionization.

The chemical and physical processes occurring after the electron impact ionization of doped helium droplets have some similarities to the processes occurring in the ISM. In both cases, a large amount of energy is inserted into a cold molecule. In the helium droplet experiment, the energy arises due to the charge transfer from the initially formed $He^+$ to a dopant. This energy is the difference between ionization potentials (IPs) of helium and a dopant. The difference in IPs of He and C atoms is about 13.32 eV. The HCH molecule formed in helium droplets before the ionization is expected to have a lower IP than that of C atoms. In H I regions of the ISM, the energy is brought to a molecule by UV photons $h\nu < 13.6$ eV. The input of a large amount of energy leads to the breaking of weak bonds and, therefore, to the formation of the most stable species. Therefore, the appearance of the "magic" peaks in the mass spectra is particularly interesting.

Ionization of helium droplets doped with pure dihydrogen results in a distribution of $H_n^+$ clusters with preferentially odd numbers of H atoms $n$ (Jaksch et al. 2008) and some well-known intensity anomalies that can be explained by an ionic $H_3^+$ core solvated by hydrogen molecules. With the addition of carbon atoms, new ions are observed. The intensity of $C_mH_n^+$ ions ($m = 1,2,3$) strongly depends on the number of the hydrogen atoms (see Figure 3). This can be easily understood, considering the covalent bonding in the formed hydrocarbon molecules.

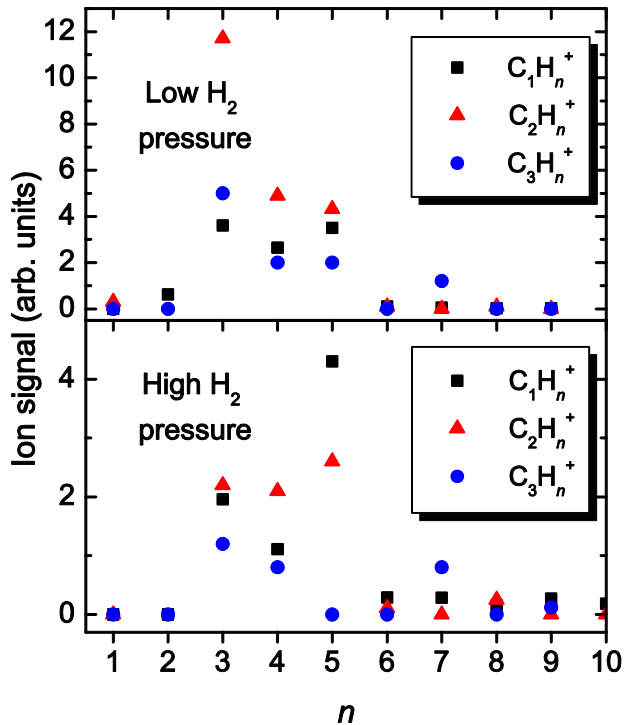

**Figure 3.** Ion signals on the masses corresponding to the products of chemical reactions between carbon atoms and dihydrogen molecules. The error bars are smaller than the size of the symbols used for plotting.

At low dihydrogen doping, the most intense "magic" peaks are caused by the $C_2H_3$ cation. After increasing the number of picked up dihydrogens, the dominant formation of the $CH_5$ cation was observed. The number of produced $CH_5^+$ cations was more than two times higher than that of $CH_3$ or $C_2H_3$ cations. In spite of the low $CH_5^+ \rightarrow CH_3^+ + H_2$ dissociation energy, which was measured to be only 40 kcal/mol (Hiraoka & Kebarle 1976), in our experiments, a high stability of $CH_5^+$ was observed. Therefore, it implies that there is a fast pathway of energy dissipation by the $CH_5$ cation. This is in line with a high rate of the radiative association found for the $CH_3^+ + H_2 \rightarrow CH_5^+ + h\nu$ reaction, which was at least one order of magnitude larger than predicted taking into account only the relaxation via vibrational transitions (Barlow et al. 1984). A previous study on the ionization of helium droplets doped with methane clusters also shows the high probability of the $CH_5^+$ formation. This demonstrates that the $CH_5^+$ formation is not dependent on the types of the precursor molecules. Therefore, in the ISM, the $CH_5^+$ can be produced not only by sequences of bottom-up reactions, but also by the top-down process when ice grains containing both carbon and hydrogen elements collide with the high energy particles.

The obtained results suggest that the relative abundances of the $CH_5^+$ among other hydrocarbon cations present in the ISM could be much higher than that considered previously (Sternberg & Dalgarno 1995). Due to the top-down formation, this could be particularly pronounced in the colder areas at extinctions $A_V > 6$.

The $CH_5^+$ was first detected in 1952, and since then it was a subject of intensive studies (Tal'roze & Lyubimova 1952, Asvany et al. 2015, Ivanov et al. 2010, White et al. 1999, Sefcik et al. 1974, Semaniak et al. 1998, Thompson et al. 2005). In spite of these thorough studies only the IR spectrum of this cation is currently available (Asvany et al. 2005, Asvany et al. 2015). At the same time, the high rate of $CH_3^+ + H_2$ radiative association suggests the presence of electronic states of $CH_5^+$ lying below or just a little above the $CH_3^+ + H_2$ dissociation limit and, consequently, the presence of absorption bands in the NIR range. The attempt to locate such states by quantum chemical computation was unsuccessful (Talbi & Saxon 1992). However, considering an extreme complexity of this cation for the computations (Ivanov et al. 2010), further studies in this direction seem to be necessary. In particular, the spectral characterization of the cation in the optical and NIR ranges should help us to locate the possible excited electronic states and to understand its influence on the spectral properties of the ISM.

## 4. CONCLUSIONS

The reactions of carbon atoms with dihydrogen were studied at low temperature in helium droplets and after the electron impact ionization of these droplets. The low-temperature reaction $C + H_2 + M \rightarrow HCH + M$ was found to be barrierless. Therefore, the reaction $C + H_2 \rightarrow CH + H$ and the reverse one $CH + H \rightarrow C + H_2$ are predicted to have no energy barrier. The $CH + H \rightarrow C + H_2$ reaction is expected to be fast in the low-temperature range of the ISM. In the case of the low C/H ratio, the ion-molecule reactions followed by electron impact ionization lead to the dominant formation of $CH_5^+$. This implies the high abundance of this cation in the ISM. The need for further spectral characterization of the $CH_5$ cation is indicated.

## ACKNOWLEDGEMENTS

The authors are grateful for the support by the by the Austrian Science Fund FWF (P26635, I978), Deutsche Forschungsgemeinschaft DFG (Contract No. JA 2107/4-1), and COST Action CM1401 "Our Astro-Chemical History."